\documentclass[showpacs,nofootinbib]{revtex4}
\usepackage{amsfonts}
\usepackage{amsmath}
\usepackage{graphicx}
\usepackage{amssymb}

\begin{document}

\title{Will the tachyonic Universe survive the Big Brake?}
\author{Zolt\'{a}n Keresztes$^{1,2}$, L\'{a}szl\'{o} \'{A}. Gergely$^{1,2}$,
Alexander Yu. Kamenshchik$^{3,4}$, Vittorio Gorini$^{5,6}$, David Polarski$%
^{7}$}
\affiliation{$^{1}$ Department of Theoretical Physics, University of Szeged, Tisza Lajos
krt 84-86, Szeged 6720, Hungary \\
$^{2}$ Department of Experimental Physics, University of Szeged, D\'{o}m T%
\'{e}r 9, Szeged 6720, Hungary\\
$^{3}$ Dipartimento di Fisica and INFN, via Irnerio 46, 40126 Bologna, Italy
\\
$^{4}$ L.D. Landau Institute for Theoretical Physics, Russian Academy of
Sciences, Kosygin street 2, 119334 Moscow, Russia\\
$^{5}$ Dipartimento di Scienze Fisiche e Mathematiche, Universit\`{a}
dell'Insubria, Via Valleggio 11, 22100 Como, Italy \\
$^{6}$ INFN, sez. di Milano, Via Celoria 16, 20133 Milano, Italy\\
$^{7}~$Laboratoire de Physique Th\'eorique et Astroparticules, CNRS,
Universit\'e Montpellier II, France}

\begin{abstract}
We investigate a Friedmann universe filled with a tachyon scalar field,
which behaves as dustlike matter in the past, while it is able to accelerate
the expansion rate of the universe at late times. The comparison with type
Ia supernovae (SNIa) data allows for evolutions driving the universe into a
Big Brake. Some of the evolutions leading to a Big Brake exhibit a large
variation of the equation of state parameter at low redshifts which is
potentially observable with future data though hardly detectable with
present SNIa data. The soft Big Brake singularity occurs at finite values of
the scale factor, vanishing energy density and Hubble parameter, but
diverging deceleration and infinite pressure. We show that the \textit{%
geodesics can be continued} through the Big Brake and that our model
universe will \textit{recollapse} eventually in a Big Crunch. Although the
time to the Big Brake strongly depends on the present values of the
tachyonic field and of its time derivative, the time from the Big Brake to the
Big Crunch represents a kind of \textit{invariant timescale} for all field
parameters allowed by SNIa.
\end{abstract}

\maketitle

\section{Introduction}

Dark energy (DE) models aim to explain the accelerated expansion rate of the
universe at late times. This phenomenon was originally discovered using
supernovae Ia (SNIa) data \cite{cosm} and has since then been confirmed by
many other observations (see e.g. \cite{Tsuji} and references therein).
Still, the nature of dark energy, and the precise physical mechanism
producing the accelerated expansion remains to date an outstanding mystery
for cosmologists and for theoretical physicists.

The $\Lambda $CDM model, based on a cosmological constant and cold dark
matter, appears to be in good agreement with most of the present
observational data on large cosmological scales. However, this model has
well known theoretical problems \cite{dark} and it also encounters
difficulties in explaining some of the data on the scales of structures and
even on very large scales, like peculiar flows (see e.g. \cite{P08}).

Alternatives to the $\Lambda $CDM model are dark energy models with a time
varying equation of state (EoS) parameter $w$ \cite{dark}, and these are not
yet excluded by data. In this context, many scalar field models have been
considered, either minimally coupled with a standard kinetic term, or more
complicated ones, e.g. Dirac-Born-Infeld (DBI) models with kinetic terms
involving a square root \cite{BI}. Interest in DBI type models was revived
in the framework of string theory, where the respective scalar fields are
called tachyons \cite{tachyons},\cite{FKS02}. These models are possible dark
energy candidates, as they can be interpreted as perfect fluids with a
sufficiently negative pressure in order to produce the late-time accelerated
expansion.

There is a large arbitrariness in the choice of the potential for tachyonic
cosmological models. In Ref \cite{we-tach} (to be referred henceforth as
\textbf{I})\ a specific tachyon potential, containing trigonometric
functions, was considered. This model turns out to be surprisingly rich, in
that it admits a large variety of cosmological evolutions depending on the
choice of initial conditions. Thus, in \textbf{I} two interesting properties
were found. First, for positive values of the model parameter $k$, the sign
of the pressure can change during evolution. Second, while under certain
initial conditions the universe will expand indefinitely towards a de Sitter
attractor, under different initial conditions, after a long period of
accelerated expansion the pressure becomes positive and the acceleration
turns into deceleration. Accordingly, the tachyon field will drive the
universe towards a new type of cosmological singularity, the Big Brake,
characterized by a sudden stop of the cosmic expansion. At this singularity,
the universe has a finite size, a vanishing Hubble parameter and an infinite
negative acceleration. In contrast to this dynamical picture, similar
evolutions were analyzed earlier from a purely kinematical standpoint and
named sudden future singularities \cite{sudden}. As already stressed earlier
for some tachyon models \cite{FKS02}, the addition of dust is a non-trivial
problem as is also the behavior of the model at the level of perturbations.
So our model cannot be viewed yet as a fully viable cosmological scenario
but rather as a toy model that could lead to a viable one after suitable
improvements. In a recent paper \cite{tach-obs} (to be referred henceforth
as \textbf{II}) we have confronted our tachyon cosmological model with SNIa
data \cite{SN2007} (see also \cite{tach-Invisible}). The strategy was the
following: for fixed values of the model parameter $k$, we scanned the pairs
of present values of the tachyon field and of its time derivative (points in
phase space) and we propagated them backwards in time, comparing the
corresponding luminosity distance - redshift curves with the observational
data from SNIa. Then, those pairs of values which appeared to be compatible
with the data, were chosen as initial conditions for the future cosmological
evolution. Though the constraints imposed by the data were severe, both
evolutions took place: one very similar to $\Lambda $CDM and ending in an
exponential (de Sitter) expansion; another with a tachyonic crossing where
the pressure turns positive from negative, ending in a Big Brake. It was
found that a larger value of the model parameter $k$ enhances the
probability to evolve into a Big Brake. For a set of initial conditions
favored by the SNIa data, we have also computed in \textbf{II} the time to
the tachyonic crossing, and the Big Brake, respectively. These time-scales
were found to be comparable with the present age of the Universe.

The purpose of the present paper is twofold. First we propose to shed more
light on the evolution of the tachyon
field in the distant and in the more recent past; 
and second, to explore in detail what happens when the universe
reaches the Big Brake. As this singularity is a soft one, with only the
second derivative of the scale factor diverging, it is expected that it may
be possible for geodesic observers to cross the singularity. Indeed, the
traversability of a rather generic class of sudden future singularities by
causal geodesics was put in evidence in \cite{FJL}. Strings can also pass
through \cite{BD}.

In Section II we consider the late-time evolution of the tachyon field, its
energy density $\rho $, pressure $p$ and EoS parameter (barotropic index) $w$
defined as $p\equiv w\rho $. In particular, we investigate whether some
observable signature today may point towards a Big Brake singularity in the
future.

In Section III we discuss the Big Brake singularity, both in terms of
curvature characteristics and by analyzing the geodesic deviation equation.
In Section IV we discuss what happens to the tachyon universe after the Big
Brake. Finally, we summarize our results with some comments in the
Concluding Remarks.

\section{Tachyon scalar field cosmology}

First, we briefly give the basic equations of tachyon cosmology. The
Lagrangian of a tachyon field is
\begin{equation}
L=-V(T)\sqrt{1-g^{\mu \nu }T_{,\mu }T_{,\nu }},  \label{L}
\end{equation}%
where $V(T)$ is some suitable tachyon potential. A homogeneous tachyon field
$T(t)$ in a Friedmann-Lema\^{\i}tre-Robertson-Walker (FLRW) universe with
metric
\begin{equation}
ds^{2}=-dt^{2}+a^{2}(t)\left[ dr^{2}+r^{2}\left( d\theta ^{2}+\sin
^{2}\theta d\varphi ^{2}\right) \right] ~,  \label{FLRW}
\end{equation}%
can be thought of as an ideal (isotropic) comoving perfect fluid with energy
density $\rho $ given by
\begin{equation}
\rho =\frac{V(T)}{\sqrt{1-\dot{T}^{2}}}~,  \label{rho}
\end{equation}%
and pressure $p$ given by
\begin{equation}
p=-V(T)\sqrt{1-\dot{T}^{2}}~,  \label{p}
\end{equation}%
where a dot denotes the derivative with respect to cosmic time $t$. The
Friedmann equation is then
\begin{equation}
H^{2}=\frac{8\pi G}{3}\frac{V(T)}{\sqrt{1-\dot{T}^{2}}},  \label{F1}
\end{equation}%
while the equation of motion for the tachyon field $T$ reads
\begin{equation}
\frac{\dot{s}}{1-s^{2}}+3Hs+\frac{V_{,T}}{V}=0~,  \label{KG}
\end{equation}%
where
\begin{equation}
s \equiv \dot{T}.  \label{s-def}
\end{equation}
Here we consider the following potential \cite{we-tach}
\begin{equation}
V(T)=\frac{\Lambda }{\sin ^{2}\left[ \frac{3}{2}{\sqrt{\Lambda \,(1+k)}\ T}%
\right] }\sqrt{1-(1+k)\cos ^{2}\left[ \frac{3}{2}{\sqrt{\Lambda \,(1+k)}\,T}%
\right] }\ ,  \label{VTfixed}
\end{equation}%
where $k$ and $\Lambda $ are free model parameters.

>From the present values $T_0$ and $s_0$ of the phase space variables 
$T$ and of its time derivative $s=\dot{T}$ we found convenient in
\textbf{II} to introduce the parameters $y_{0}$ and $x_{0}$ (denoted $w_{0}$
in \textbf{II}) defined as
\begin{equation}
y_{0}=\cos \left[ \frac{3}{2}{\sqrt{\Lambda \,(1+k)}\,T_{0}}\right] ~,
\end{equation}%
\begin{equation}
x_{0}=\frac{1}{1+s_{0}^{2}}~.
\end{equation}

The Hubble parameter $H$ as a function of the redshift $z$ is expressed as
\begin{equation}
H^{2}(z)=H_{0}^{2}~\Omega _{T,0}~\frac{\rho (z)}{\rho _{0}}~,
\end{equation}%
with $\Omega _{T,0}\equiv \frac{\rho _{0}}{\rho _{cr,0}}$, which can be
computed in principle as follows
\begin{equation}
\frac{\rho (z)}{\rho _{0}}=\exp \left[ 3\int_{0}^{z}dz^{\prime }~\frac{%
1+w(z^{\prime })}{1+z^{\prime }}\right] ~,
\end{equation}%
where $w=\frac{p}{\rho }$ can be obtained from (\ref{rho}) and (\ref{p}).

As mentioned in the Introduction, we consider here a model containing only
the tachyon field $T$. Hence, with respect to the expansion rate, the EoS
parameter $w$ of $T$ should be compared to what is usually called $w_{%
\mathrm{eff}}\equiv w_{\mathrm{DE}}\Omega _{\mathrm{DE}}$, for a universe
filled with a dark energy component and dust-like matter. In particular, for
$\Lambda $CDM we have $w_{\mathrm{eff}}=-\Omega _{\Lambda }$.

Finally it is instructive to write down the second Friedmann equation
\begin{equation}
\frac{\ddot a}{a} = -\frac{4\pi G}{3}(\rho + 3 p) ~.  \label{F2}
\end{equation}
The Big Brake corresponds to vanishing energy density $\rho$ and infinite
(positive) pressure $p$. Hence $\dot{a}=0$ from (\ref{F1}) while $\ddot{a}%
=-\infty$ from (\ref{F2}) at the Big Brake (whence the name). It is reached
for finite time and finite value of the scale factor.

\subsection{Evolution of the system}

\begin{figure}[t]
\vskip0.5cm \includegraphics[height=8cm, angle=270]{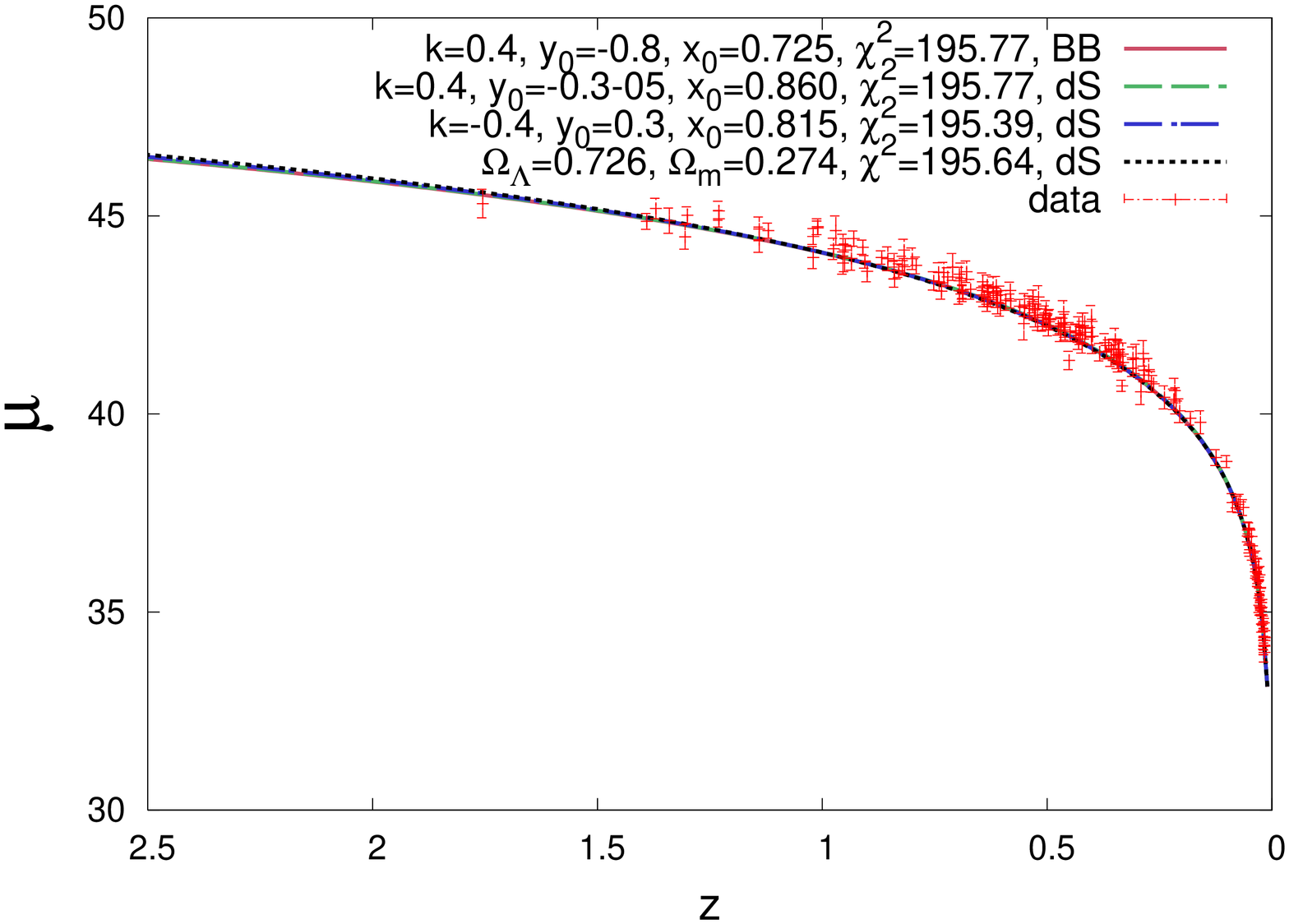} %
\includegraphics[height=8cm, angle=270]{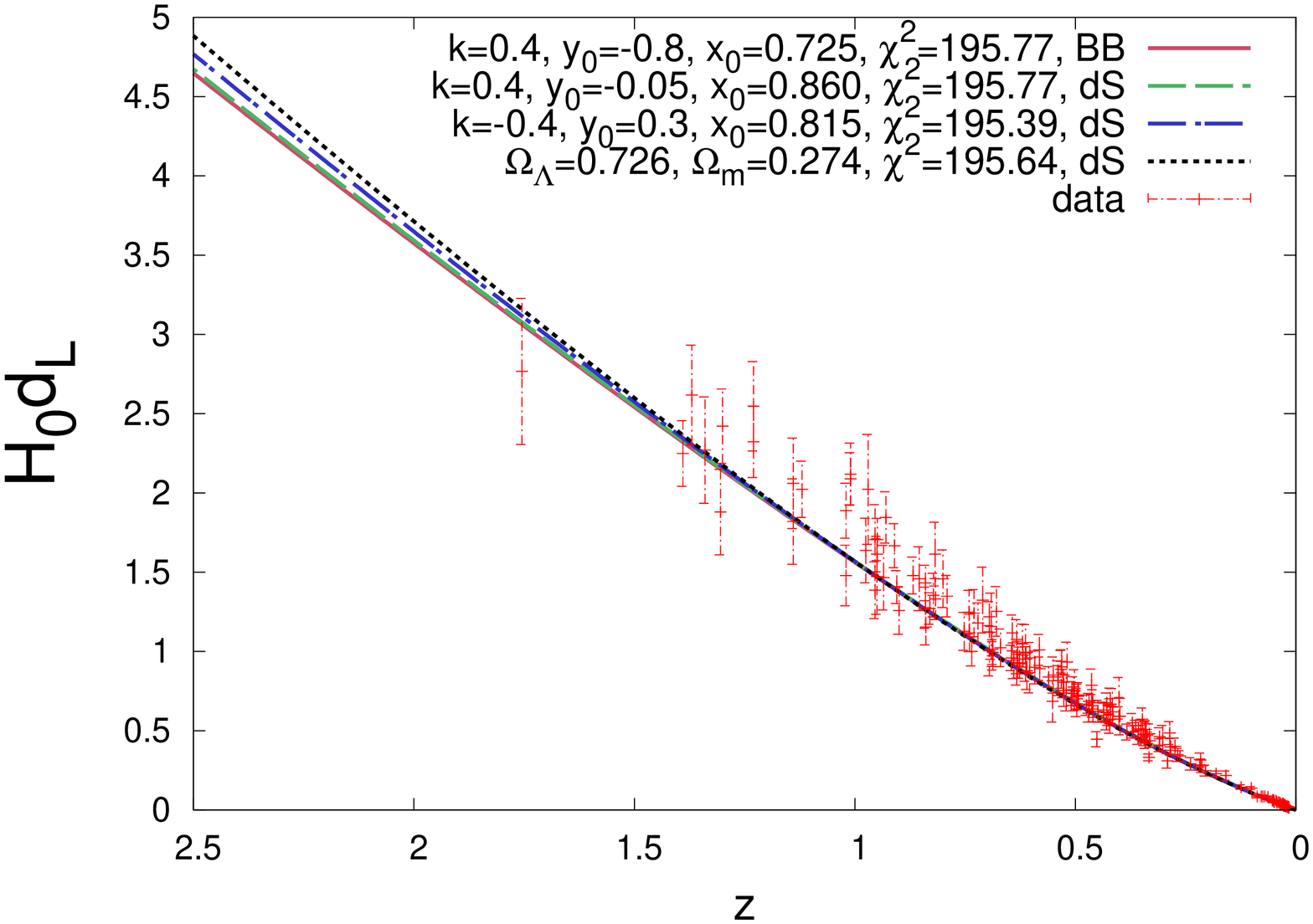}
\caption{(Color online) The distance modulus (left panel) and dimensionless
version (with $c=1$) of the luminosity distance (right panel) in recent
cosmological times. The models shown here are $\Lambda $CDM (black); the
tachyonic model with $k=-0.4$\ evolving into de Sitter (dS, blue); tachyonic
models with $k=0.4$ evolving into de Sitter (green) and into a Big Brake
(BB, red).\ \ The curves are all in very good agreement with the SNIa data,
as they pass close to the local minimum of the (respective regions of the) 1-%
$\protect\sigma $ domains selected by supernovae (see Figs 2 and 3 of
\textbf{II}.)}
\label{Fig1}
\end{figure}

We consider the evolution of the trajectories of the model compatible with
the supernovae data \cite{SN2007} at the 1-$\sigma $ level. To this purpose
we first display on Fig \ref{Fig1} the behavior of the distance modulus and
of the luminosity distance as functions of redshift $z$, for 4 different
models, all of them fitting within 1-$\sigma $ accuracy the data (actually,
the curves have the best fits in their respective model classes). We recall
that the distance modulus $\mu $ is defined as
\begin{equation}
\mu =5\log _{10}\frac{d_{L}\left( z\right) }{\text{Mpc}}+25~,
\end{equation}%
where $d_{L}(z)$ is the luminosity distance which, for a flat Friedmann
universe, is given by ($c=1$)
\begin{equation}
d_{L}\left( z\right) =\left( 1+z\right) \int_{0}^{z}\frac{dz^{\ast }}{%
H\left( z^{\ast }\right) }~.
\end{equation}%

\begin{figure}[t]
\vskip0.5cm \includegraphics[height=8cm, angle=270]{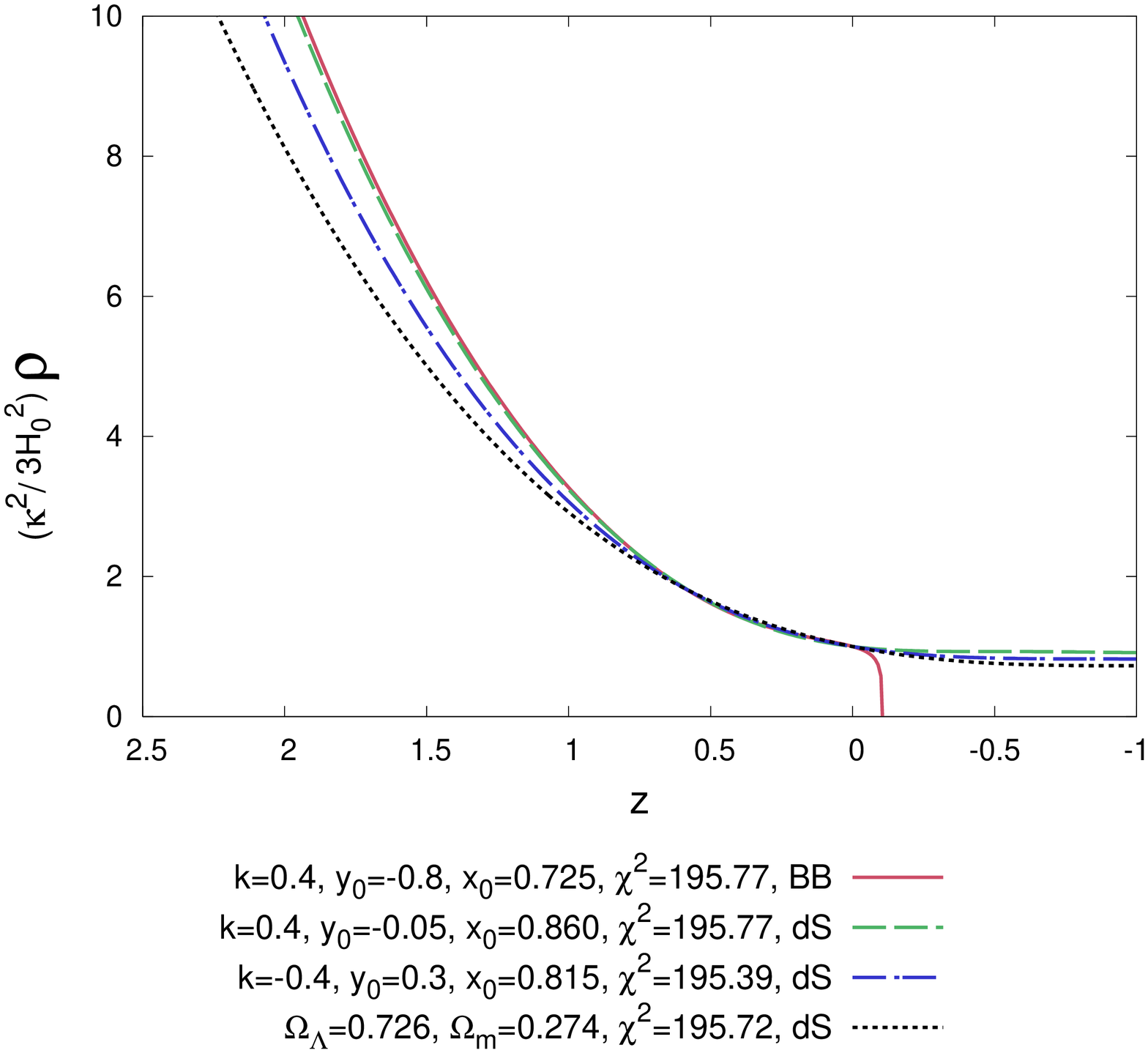} %
\includegraphics[height=8cm, angle=270]{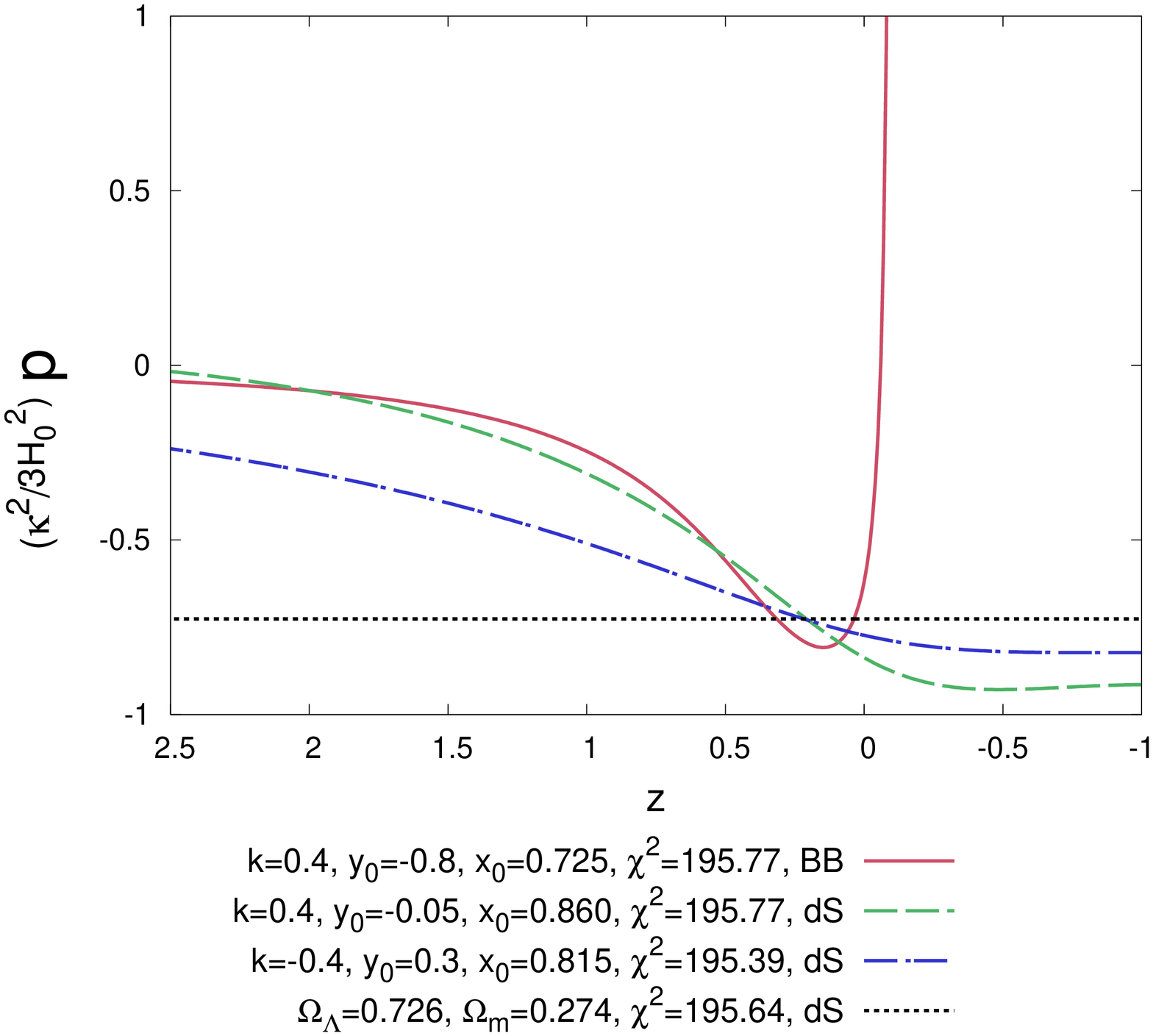} \vskip0.5cm
\caption{(Color online) The evolution in the recent past of the universe and
in the future of the normalized energy density $\frac{\protect\rho }{\protect%
\rho _{cr,0}}$ (left panel) and of the normalized pressure $\frac{p}{\protect%
\rho _{cr,0}}$ (right panel) is shown for four models. One of the models
leads to a Big Brake singularity in the future while the other three models
shown tend asymptotically to a de Sitter space.}
\label{Fig2}
\end{figure}

\begin{figure}[h]
\vskip1cm \includegraphics[height=8cm, angle=270]{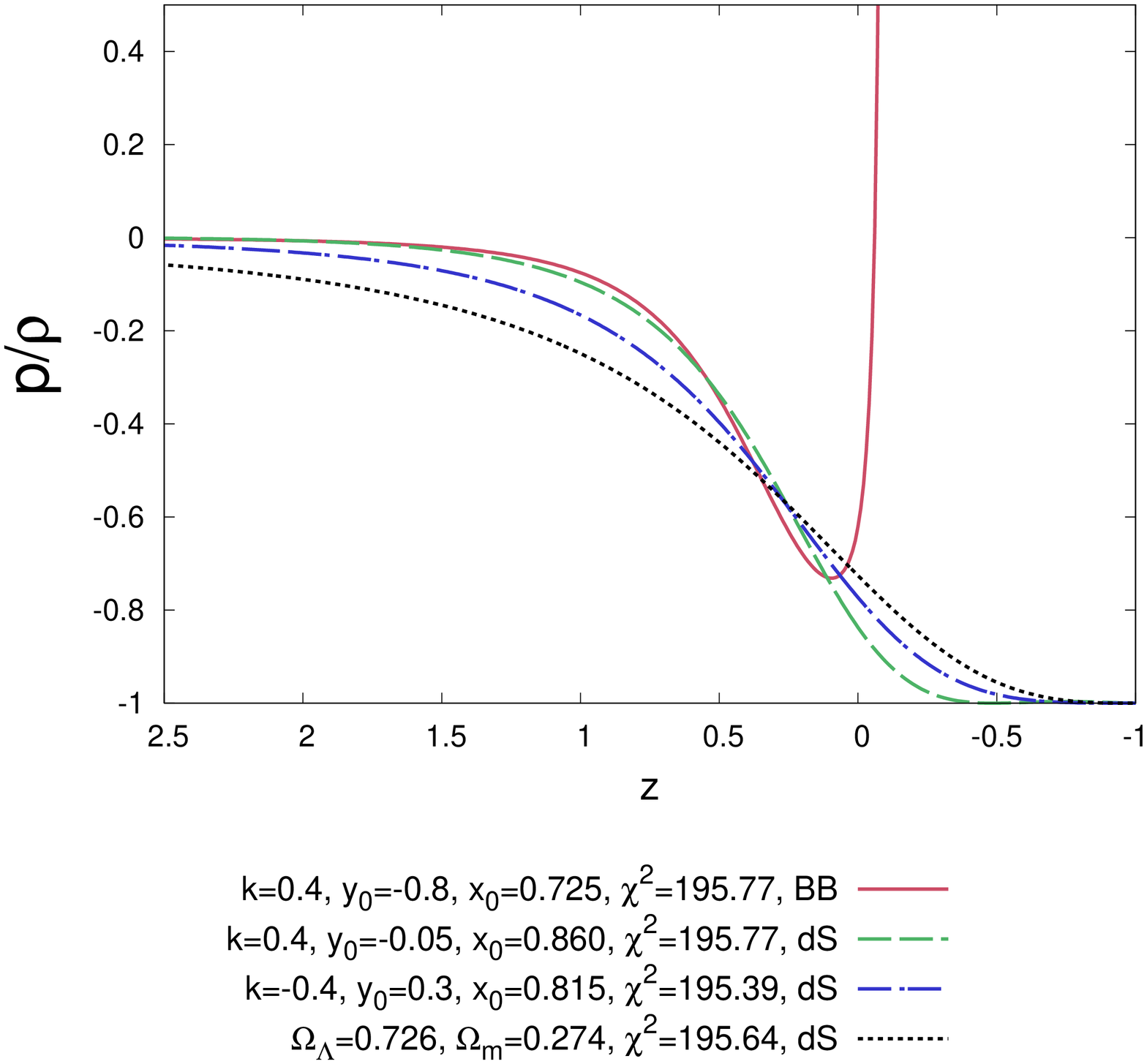} %
\includegraphics[height=8cm, angle=270]{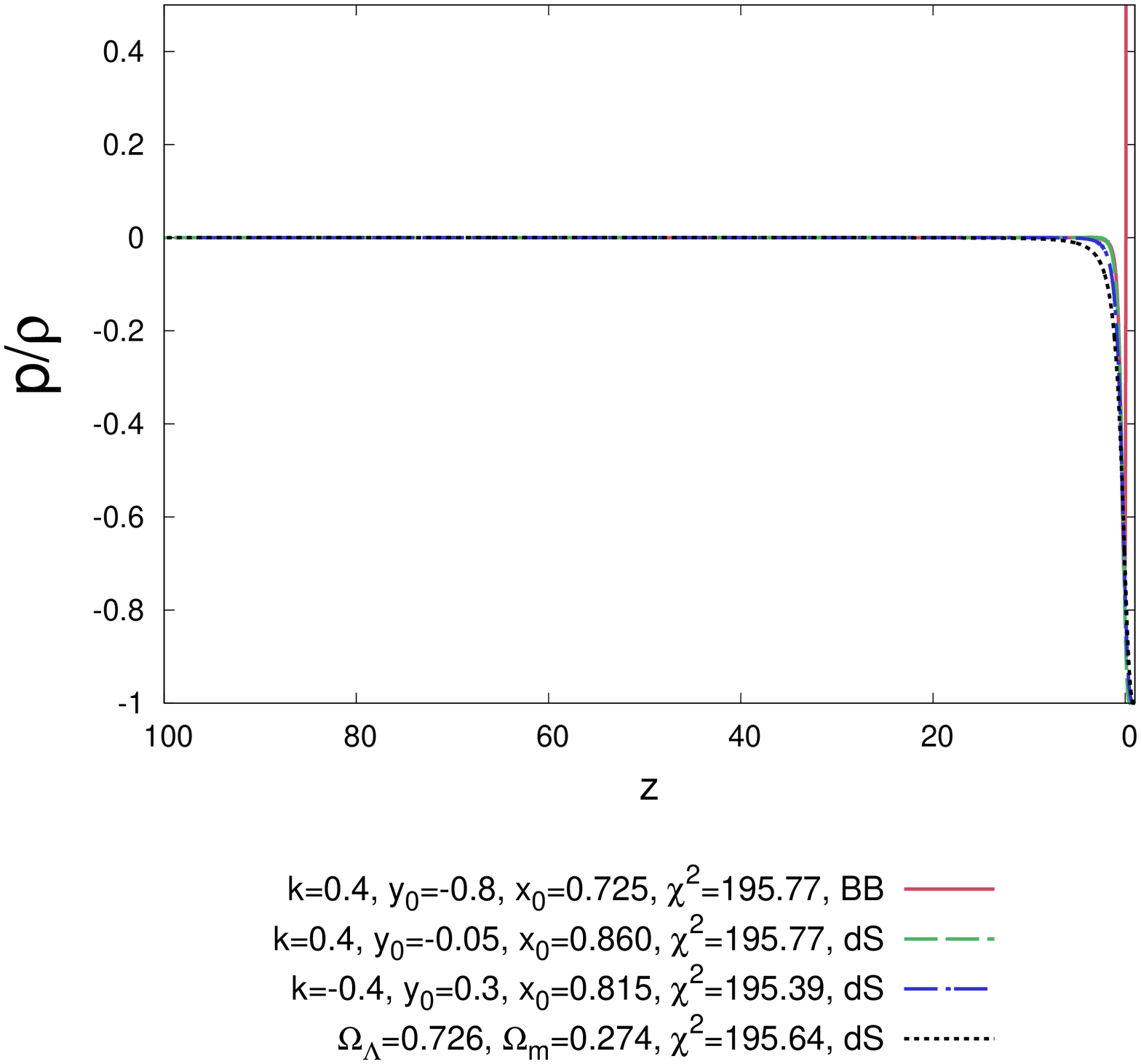} \vskip0.5cm
\caption{(Color online) The evolution of the EoS parameter $w$ ($w_{\mathrm{%
eff}}$ for the $\Lambda $CDM model) in the recent past and in the future is
shown for the same four models, at late times from $z=2.5$ to $z=-1$ (left
panel); and in the distant past from $z=100$ on (right panel). It is seen
that the tachyon field behaves essentially as dustlike matter at high
redshifts.}
\label{Fig3}
\end{figure}

For these specific evolutions we also show the normalized dimensionless
energy density $\frac{\rho }{\rho _{cr,0}}$, pressure $\frac{p}{\rho _{cr,0}}%
=w\frac{\rho }{\rho _{cr,0}}$ and EoS parameter $w$ (being $w_{\mathrm{eff}}$
for the $\Lambda $CDM model) as function of the redshift (both in the past
and in the future) on Figs \ref{Fig2}-\ref{Fig3}.

Four curves appear on each graph on Figs \ref{Fig1}-\ref{Fig3}. The black
curves refer to the $\Lambda $CDM model (with value of $\Omega _{\Lambda ,0}$
taken from WMAP analysis \cite{WMAP}). The dark matter component evolves
here as $(1+z)^{3}$. The other three curves are for the toy model containing
only the tachyon field. The three curves differ in the model parameter $k$,
and in the initial data $x_{0}$, $y_{0}$. All three curves pass close to the
local minimum of the respective 1-$\sigma $ domains selected by type Ia
supernovae. They are characterized by the model parameter $k=-0.4$ (blue)
and $k=0.4$, respectively. For the latter we have picked up both types of
allowed evolutions, one going into de Sitter (green), the other ending in a
Big Brake (red).
Note also that nowadays we have $w_{0}\equiv w(z=0)\in \lbrack -0.8,-0.6]$
in all the evolutions displayed. These values are similar to $\Lambda $CDM
where $w_{\mathrm{eff,0}}=-\Omega _{\Lambda ,0}\approx -0.74$. As expected
from viable cosmological evolutions, the parameter $w$ approaches zero in
the past, corresponding to dust-like behavior at early times. For all
parameters, the pressure remains very slightly negative for the whole
positive $z$ range plotted, thus $w$ stays below zero (although it gets very
close to it). Under the assumption that this behavior is unchanged when dust
is added, we expect to have a model where dark energy (here the tachyon
field) remains nonnegligible in the early stages of the universe, with $%
\Omega _{T}$ remaining roughly constant (actually slightly increasing)
during the matter era before it would start dominating at late times. A
similar behavior can also be achieved in some scalar-tensor DE models, see
e.g. \cite{GP08}.

\subsection{Observable signature of the Big Brake?}

\begin{figure}[t]
\includegraphics[height=8cm, angle=270]{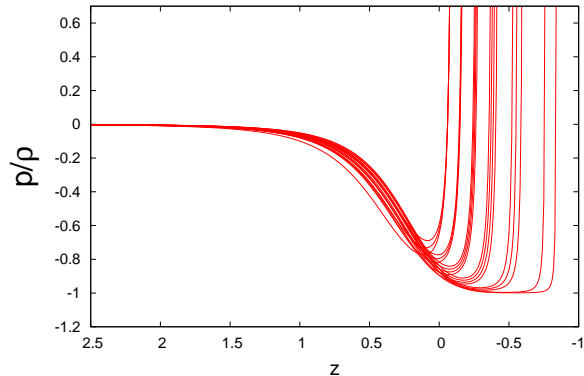} \vskip0.5cm
\caption{(Color online) The evolution of the EoS parameter $w$ in the recent
past and in the future for model parameters leading to a Big Brake and in
the 1-$\protect\sigma $ domain of supernova data. All evolutions have a dip
(when, as $z$ decreases, the decrease in $w$\ turns into an increase), some
of them already in the recent past.}
\label{Fig4}
\end{figure}

Though the Big Brake is a singularity taking place in the future, it is
interesting to investigate whether models leading to a Big Brake exhibit
some signature in our \textit{present-day} universe. As one can see from Fig %
\ref{Fig4}, this is indeed the case for some of the universes with a Big
Brake in the future which have a characteristic behavior of the EoS
parameter $w$ in our past: a \textquotedblleft dip\textquotedblright\ at low
redshifts. This happens when the Big Brake is not too far in the future, in
other words when the final redshift is substantially larger than $-1$. It is
then interesting to investigate whether such a behavior can be detected.
Actually large variations of $w$ at low redshifts can easily hide in
luminosity-distance curves. Therefore, we expect that it is very difficult
to observe this peculiar behavior with \textit{present} SNIa data. Let us
consider why this is the case in more details.

In our model we have only one component, the tachyon field, so what we call
here $w$ is actually what should be called $w_{eff}$ when more fluids are
present. However, it would be easy to extend our analysis to the case when a
dust-like component is also present. The EoS parameter $w$ has a minimum at
some small redshift $z_{min}$ with $z_{min}\sim 0.2$ in the most favorable
cases. In principle this can be tested and it is straightforward to derive
the following equality
\begin{equation}
(1+z_{min}) \frac{d \ln f}{dz}\Big|_{z_{min}} = -1~,  \label{wmin}
\end{equation}
as well as the inequalities
\begin{eqnarray}
(1+z) \frac{d \ln f}{dz} &<& -1 ~~~~~~~~~~~~~~~~~~~~~~~~~z<z_{min}
\label{dw1} \\
(1+z) \frac{d \ln f}{dz} &>& -1 ~~~~~~~~~~~~~~~~~~~~~~~~~z>z_{min}~,
\label{dw2}
\end{eqnarray}
where the quantity $f$ is easily expressed in terms of the
luminosity-distance, viz.
\begin{equation}
f = -2 \frac{d}{dz} \left[ \ln \left( \frac{d_L}{1+z} \right)^{\prime }%
\right] ~,
\end{equation}
a prime standing for the derivative with respect to $z$.

Hence it is seen that the (in)equalities (\ref{wmin}),(\ref{dw1}),(\ref{dw2}%
) imply a condition on the third derivative of the luminosity distance $%
d_{L} $! It is clear that even small uncertainties on $d_{L}(z)$ can lead to
large uncertainties on its derivatives thereby rendering the (in)equalities (%
\ref{wmin}),(\ref{dw1}),(\ref{dw2}) ineffective. In addition, as we have to
apply these conditions on low redshifts, these involve third order
contributions in $z$ to $d_{L}(z)$ which are inevitably small. Obviously,
observational uncertainties are presently too high in order to make use of (%
\ref{wmin}),(\ref{dw1}),(\ref{dw2}) with existing SNIa data.

Still, we have tried to see whether a standard $\chi ^{2}$ analysis making
use only of SNIa data in the range $0\leq z\leq z_{min}$ could differentiate
models with and without dip. This means that we assume a priori that the
model with the dip at $z_{min}$ is the correct one and that we investigate
whether a simple statistical analysis of this relevant part of the data
could hint at its presence. As could be expected, even in this case we find
no statistical evidence for the detection of a dip with the present SNIa
data. Note that even the Constitution dataset \cite{CfA09} contains only 147
SN data at redshifts $z\leq 0.2$. Though more refined statistical tools
should clearly be used (see e.g. \cite{HPZ09}), it is quite obvious that the
present data do not allow for an unambiguous detection. Note that in some
models a characteristic smoother variation of w can take place on a larger
range of redshifts (see e.g. \cite{DHRS09}) which should be easier to
detect. Models involving a very large variation of $w$ at extremely low
redshifts $z\ll 1$ were considered in \cite{MHH09} and it was found that
they could escape all high precision measurements. In our case, variations,
though not as large, are located at higher redshifts. We conjecture that
future SNIa surveys, like e.g. the Large Synoptic Survey Telescope (LSST)
and Wide Field InfraRed Survey Telescope (WFIRST) containing many more
supernovae and reducing significantly the systematic and statistical errors
could allow for such a detection. Future surveys like Euclid involving
weak-lensing are also promising in this respect. We believe this could be an
interesting scientific goal for these surveys especially if peculiar models
with a large variation of their EoS parameter at low, but not too low,
redshifts, like some of our Big Brake models with a dip, are in good
agreement with observations and are theoretically motivated candidates.

\section{The Big Brake singularity}

\subsection{Curvature}

The 3-spaces with $t=$const have vanishing Riemann curvature
\begin{equation}
^{\left( 3\right) }\mathcal{R}_{abcd}=0\,\ .
\end{equation}%
The 4-dimensional Riemann curvature tensor has therefore but few
nonvanishing independent components:%
\begin{eqnarray}
R_{trtr} &=&-\ddot{a}a\,\ ,  \notag \\
R_{t\varphi t\varphi } &=&R_{t\theta t\theta }\sin ^{2}\theta =-\ddot{a}%
ar^{2}\sin ^{2}\theta \,\ ,  \notag \\
R_{r\varphi r\varphi } &=&R_{r\theta r\theta }\sin ^{2}\theta =\dot{a}%
^{2}a^{2}r^{2}\sin ^{2}\theta \,\ ,\,\   \notag \\
R_{\theta \varphi \theta \varphi } &=&\dot{a}^{2}a^{2}r^{4}\sin ^{2}\theta \
,
\end{eqnarray}%
and the corresponding components arising from symmetry. Remarkably, all
components which diverge at the Big Brake are of the type $R_{tata}$.
Therefore, the singularity arises in the mixed spatio-temporal components.

\subsection{Geodesic deviation}

The geodesic deviation equation along the integral curves of $u=\partial
/\partial t$ (which are geodesics with affine parameter $t$) is%
\begin{equation}
\dot{u}^{a}=-R_{\ \ cbd}^{a}\eta ^{b}u^{c}u^{d}\ ,
\end{equation}%
where $\eta ^{b}$ is the deviation vector separating neighboring geodesics,
chosen to satisfy $\eta ^{b}u_{b}=0$. In the coordinate system (\ref{FLRW})
we obtain
\begin{equation}
\dot{u}^{a}=-R_{\ \ tbt}^{a}\eta ^{b}\propto \ddot{a}\ ,
\end{equation}%
which at the Big Brake diverges as $-\infty $. Therefore, when approaching
the Big Brake, the tidal forces manifest themselves as an infinite braking
force stopping the further increase of the separation of geodesics. This can
be also seen from the behavior of the velocity
\begin{equation}
v^{a}=u^{b}\nabla _{b}\eta ^{a}\propto H~,
\end{equation}%
which at the Big Brake vanishes. Immediately after, the negative
acceleration will cause the geodesics to approach each other. Therefore a
contraction phase will follow: everything that has reached the Big Brake
will bounce back.

We conclude the section with the remark that despite the singularity of the
geometry (the second derivative $\ddot{a}$ of the scale factor diverges at $%
t=t_{BB}$), its soft character ($\dot{a}$ stays regular) assures that a
continuation of the evolution is still possible in the following sense. We
indeed need to know $\ddot{a}$ to follow the evolution of the spacetime but
we only need to know $\dot{a}$ to follow the evolution of free particles.
This means that despite not being able to continue the evolution of the
geometry in a direct way, we can univocally continue the individual world
lines of freely falling test particles (geodesics), each of these being
perfectly regular at $t=t_{BB}$. The singularity is not experienced by any
individual freely falling particle, but makes itself felt only through the
equation of geodesic deviation, which at $t=t_{BB}$ indicates that the
expansion of the geodesic congruence turns negative from positive.

Once the particles have gone through the Big Brake, we can again start to
evolve the geometry itself, thus following the further evolution of the
universe beyond the singularity. As will be shown in the following section,
the tachyonic universe will evolve along similar trajectories to those
starting from a Big Bang, but in the opposite direction (with $s\rightarrow
-s$),$\,$\ arriving therefore into a Big Crunch.

\section{From the Big Brake to the Big Crunch}

\subsection{How the universe crosses the Big Brake singularity}

To understand how the crossing of the Big Brake singularity takes place, and
what is going on after the crossing, it is convenient to refer to the phase
portrait of the model at some positive value of the parameter $k$. This
portrait was drawn in Paper \textbf{I} and we reproduce it here, see Fig. 5.
\begin{figure}[t]
\includegraphics[height=7cm]{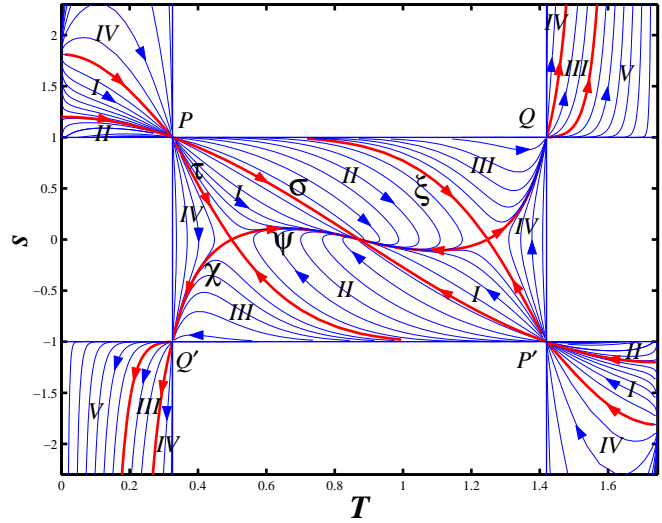}
\caption{(Color online) Phase portrait evolution for $k > 0$ ($k = 0.44$)}
\label{Fig5}
\end{figure}
The accessible phase space of the model consists of a rectangle $(T_3 \leq T
\leq T_4, -1 \leq s \leq 1)$ and four stripes. The values
\begin{eqnarray}
&&T_3 = \frac{2}{3\sqrt{(1+k)\Lambda}}\mathrm{arccos}\frac{1}{\sqrt{1+k}},
\notag \\
&&T_4 = \frac{2}{3\sqrt{(1+k)\Lambda}}\left(\pi - \mathrm{arccos}\frac{1}{%
\sqrt{1+k}}\right)  \label{T3}
\end{eqnarray}
are those for which the potential (\ref{VTfixed}) vanishes. Inside the
rectangle the dynamics of the system is described by the Lagrangian (\ref{L}%
) with the potential (\ref{VTfixed}), while in the strips the Lagrangian is
given by
\begin{equation}
L=W(T)\sqrt{g^{\mu \nu }T_{,\mu }T_{,\nu }-1},  \label{L1}
\end{equation}
with the potential
\begin{equation}
W(T) = \frac{\Lambda }{\sin ^{2}\left[ \frac{3}{2}{\sqrt{\Lambda \,(1+k)}\ T}%
\right] }\sqrt{(1+k)\cos ^{2}\left[ \frac{3}{2}{\sqrt{\Lambda \,(1+k)}\,T}%
\right]-1 }\ .  \label{W}
\end{equation}

Consider a trajectory entering the left lower strip through the point Q'
having coordinates ($T = T_3, s = -1$). The analysis of the equations of
motion, carried out in Paper \textbf{I}, has shown that the universe
encounters a Big Brake (BB) singularity after a finite time. This
singularity is characterized by some value of the tachyon field $T_{BB}$, of
the time $t_{BB}$ and of the value of the cosmological radius $a_{BB}$.
These values are found numerically up to normalization, as was done in
\textbf{II}.

The equation of motion (\ref{KG}) for the expanding universe in the left
lower strip can be written as
\begin{eqnarray}
&&\dot{s} = \frac{3s(s^2-1)\sqrt{\Lambda}}{\sin \frac{3\sqrt{\Lambda(1+k)}T}{%
2}} \left(\frac{(k+1)\cos^2 \frac{3\sqrt{\Lambda(1+k)}T}{2}-1}{s^2-1}%
\right)^{1/4}  \notag \\
&&-\frac{3\sqrt{\Lambda(1+k)}(s^2-1)}{2}\cot\frac{3\sqrt{\Lambda(1+k)}T}{2}
\notag \\
&&\times \frac{(k+1)\cos^2 \frac{3\sqrt{\Lambda(1+k)}T}{2} + (k-1)}{%
(k+1)\cos^2 \frac{3\sqrt{\Lambda(1+k)}T}{2} -1}.  \label{sdot0}
\end{eqnarray}
>From the analysis of this equation  we find that, approaching the Big Brake
singularity in the lower left strip of the phase diagram, the tachyon field $%
T$, its time derivative $s$, the cosmological radius $a$, its time
derivative $\dot{a}$ and the Hubble variable $H$ behave respectively as
\begin{equation}
T=T_{BB}+\left( \frac{4}{3W(T_{BB})}\right) ^{1/3}(t_{BB}-t)^{1/3},
\label{tachBB}
\end{equation}
\begin{equation}
s=-\left( \frac{4}{81W(T_{BB})}\right) ^{1/3}(t_{BB}-t)^{-2/3},  \label{sBB}
\end{equation}%
\begin{equation}
a=a_{BB}-\frac{3}{4}a_{BB}\left( \frac{9W^{2}(T_{BB})}{2}\right)
^{1/3}(t_{BB}-t)^{4/3},  \label{cosmradBB}
\end{equation}%
\begin{equation}
\dot{a}=a_{BB}\left( \frac{9W^{2}(T_{BB})}{2}\right) ^{1/3}(t_{BB}-t)^{1/3},
\label{cosmradderBB}
\end{equation}%
\begin{equation}
H=\left( \frac{9W^{2}(T_{BB})}{2}\right) ^{1/3}(t_{BB}-t)^{1/3}.
\label{HubbleBB}
\end{equation}%
To arrive to formulae (\ref{tachBB})--(\ref{HubbleBB}) we have used the
following strategy. Assume that in the neighborhood of the Big Brake
singularity the tachyon field behaves as
\begin{equation}
T = T_{BB} + A (t_{BB}-t)^{\alpha},  \label{asymp}
\end{equation}
where $A$ and $\alpha$ are some real parameters to be determined. Then, $s$
behaves as
\begin{equation}
s = -\alpha A(t_{BB}-t)^{\alpha-1},  \label{asymp1}
\end{equation}
while its time derivative is
\begin{equation}
\dot{s} = \alpha(\alpha-1)A(t_{BB}-t)^{\alpha-2}.  \label{asymp2}
\end{equation}
A simple calculation shows that the first ``friction'' term, proportional to
the Hubble variable in the right-hand side of Eq. (\ref{sdot0}), has the
behavior
\begin{equation}
s^{5/2} \sim (t_{BB}-t)^{5(\alpha-1)/2},
\end{equation}
which is stronger than the corresponding behavior of the second potential
term in the right-hand side of Eq. (\ref{sdot0}) which is
\begin{equation}
s^2 \sim (t_{BB}-t)^{2(\alpha-1)}.
\end{equation}
This means that the term $\dot{s}$ in the left-hind side of Eq. (\ref{sdot0}%
) should have the same asymptotic as the friction term in the right-hand
side of the same equation and, hence
\begin{equation}
\alpha - 2 = \frac52(\alpha-1),
\end{equation}
which gives immediately
\begin{equation}
\alpha = \frac13.
\end{equation}
Comparing the coefficients of the leading terms in Eq. (\ref{sdot0}) we find
that
\begin{equation}
A = \left( \frac{4}{3W(T_{BB})}\right).
\end{equation}
Thus, we arrive at Eq. (\ref{tachBB}). Eq. (\ref{sBB}) follows right away.
Using the Friedmann equation we obtain the value of the Hubble parameter
(Eq. (\ref{HubbleBB})), which, in turn, gives formulae (\ref{cosmradBB}) and
(\ref{cosmradderBB}) for the cosmological radius and for its time derivative.

The expressions (\ref{tachBB})--(\ref{HubbleBB}) can be continued in the
region where $t>t_{BB}$, which amounts to crossing the Big Brake
singularity. Only the expression for $s$ is singular at $t=t_{BB}$, but this
singularity is integrable and not dangerous.

Upon reaching the Big Brake, it is impossible for the system to stop there
because  the infinite deceleration  eventually leads to the decrease
of the scale factor. This is because after the Big Brake crossing the time
derivative of the cosmological radius (\ref{cosmradderBB}) and of the Hubble
variable (\ref{HubbleBB}) change their signs. The expansion is then followed
by a contraction.

Corresponding to given initial conditions, we can find numerically the
values of $T_{BB}$, $t_{BB}$ and $a_{BB}$ (see Paper \textbf{II}). Then, in
order to see what happens after the Big Brake crossing, we can choose as
initial conditions for the \textquotedblleft after-Big-Brake-contraction
phase\textquotedblright\ some value $t=t_{BB}+\varepsilon $ and the
corresponding expressions for $T,s,H,a$ and $\dot{a}$ following from
relations (\ref{tachBB})--(\ref{HubbleBB}), and integrate numerically the
equations of motion, thus arriving eventually to a Big Crunch singularity.

\subsection{What is going on after the Big Brake crossing ?}

After the Big Brake crossing the universe has a negative value of the
variable $s$, less than $-1$. This means that its evolution should end in a
finite period of time. Remember that the universe is now squeezing. The
equation of motion for $s$ then looks as follows:
\begin{eqnarray}
&&\dot{s} = -\frac{3s(s^2-1)\sqrt{\Lambda}}{\sin \frac{3\sqrt{\Lambda(1+k)}T%
}{2}} \left(\frac{(k+1)\cos^2 \frac{3\sqrt{\Lambda(1+k)}T}{2}-1}{s^2-1}%
\right)^{1/4}  \notag \\
&&-\frac{3\sqrt{\Lambda(1+k)}(s^2-1)}{2}\cot\frac{3\sqrt{\Lambda(1+k)}T}{2}
\notag \\
&&\times \frac{(k+1)\cos^2 \frac{3\sqrt{\Lambda(1+k)}T}{2} + (k-1)}{%
(k+1)\cos^2 \frac{3\sqrt{\Lambda(1+k)}T}{2} -1}.  \label{sdot}
\end{eqnarray}

In principle, the evolution of the universe can  either end at the vertical
line $T=0$ at some value of $s$ or at the horizontal line $s=-1$ at some
value of $T$. One can find the corresponding points on the phase diagram by
direct analysis of the system of equations of motion. However, such an
analysis is rather cumbersome. Thus it is convenient to use some results of
the analysis of the trajectories for the expanding universe given in Paper
\textbf{I}. Begin by writing down the equation for the trajectories
describing the \textit{expanding} universe in the phase space ($T,s$),
eliminating the time parameter $t$:
\begin{eqnarray}
&&\frac{ds}{dT} = -\frac{3(1-s^2)\sqrt{\Lambda}} {\sin \frac{3\sqrt{%
\Lambda(1+k)}T}{2}} \left(\frac{1-(k+1)\cos^2 \frac{3\sqrt{\Lambda(1+k)}T}{2}%
}{1-s^2}\right)^{1/4} +  \notag \\
&&-\frac{3\sqrt{\Lambda(1+k)}}{2} \frac{1-s^2}{s}\cot \left(\frac{3\sqrt{%
\Lambda(1+k)}T}{2}\right) \frac{(k+1)\cos^2\frac{3\sqrt{\Lambda(1+k)}T}{2} +
(k-1)}{1 - (k+1)\cos^2 \frac{3\sqrt{\Lambda(1+k)}T}{2}}.  \label{eq-trajec}
\end{eqnarray}
This equation is valid in both the rectangle of the phase diagram and in the
strips. In \textbf{I} we have considered it in the upper left strip and saw
there that the trajectories of the phase diagram can have their beginning
only in the points $(T=0,s=\sqrt{\frac{k+1}{k}})$, $(T=0,s=\sqrt{k+1})$ or $%
(T=T_{\ast },s=1),0<T_{\ast }<T_{3}$. Now, it is easy to see from Eq. (\ref%
{KG}) that the simultaneous change of sign of the Hubble parameter $\dot{a}/a
$ and of the time derivative of the tachyon field leaves this equation
invariant. This means that the trajectories describing the expansion from
the Big Bang singularity in the upper left strip are symmetrical reflections
with respect to the axis $s=0$ of the trajectories describing the
contraction towards the Big Crunch singularity in the lower left strip.
Thus, equation (\ref{eq-trajec}) written above describes also the
trajectories of the contracting universe in the lower left strip. In turn,
this implies that all the contracting trajectories can only end at the
points $(T=0,s=-\sqrt{\frac{k+1}{k}})$, $(T=0,s=-\sqrt{k+1})$ or $(T=T_{\ast
},s=-1),0<T_{\ast }<T_{3}$.

We are now in a position to analyze the behavior of all these trajectories,
describing the contracting universe, using the results obtained for the
corresponding trajectories, born in the left upper strip, describing the
expanding universe. First, all the contracting trajectories encountering the
Big Crunch singularity at the point $(T=T_{\ast },s=-1)$, where $0 <
T_{\ast} < T_3$ and the unique trajectory ending in the point $(T=0,s=-\sqrt{%
k+1})$ enter the lower left strip from the rectangle of the phase diagram
through the corner $(T=T_{3},s=-1)$ without arriving from the Big Brake
singularity (indeed, they originate from the repelling de Sitter node).
These trajectories are the time-reversed of the corresponding expanding
trajectories, having their origin at the points $(T=T_{\ast}, s = +1)$ and
of the unique trajectory originating at the point $(T = 0, s = +\sqrt{k+1})$%
, which enter into the rectangle through the point P (see Fig. 5)  and which
do not undergo any change of the expansion regime.

Thus, the only point where the trajectories coming from the crossing of the
Big Brake singularity can end to is $(T=0, s = -\sqrt{\frac{k+1}{k}})$. Now
recall (see Paper \textbf{I}) that the trajectories born at  $(T=0,s=\sqrt{%
\frac{k+1}{k}})$ behave at the beginning of their evolution as
\begin{equation}
s \approx \sqrt{\frac{k+1}{k}} + D T^{\frac{2(1-k)}{1+k}},  \label{beginning}
\end{equation}
where the parameter $D$ can take any real value. Among these, those with a
sufficiently large positive value of $D$, say, $D_{sep} < D$ grow without
limit and do not achieve a maximal value of the variable $s$. Instead, they
approach asymptotically to the vertical line $T=T_{BB}, s\rightarrow +\infty
$, thus encountering a Big Brake shortly after the Big Bang (such a
possibility was overlooked in \textbf{I}). Instead, the trajectories, for
which $D < D_{sep}$ achieve some maximal value of the variable $s$ after
which they turn down and enter the rectangle of the phase diagram through P.
Such trajectories were described in detail in Paper \textbf{I}. The
trajectory characterized by the critical value of the parameter $D = D_{sep}$
plays the role of separatrix between these two sets of the evolutions.

The trajectories approaching the Big Crunch at the point $(T=0, s = -\sqrt{%
\frac{k+1}{k}})$  behave as
\begin{equation}
s \approx -\sqrt{\frac{k+1}{k}} - D T^{\frac{2(1-k)}{1+k}}.  \label{ending}
\end{equation}
Those with $D > D_{sep}$ are the evolutions which underwent the Big Brake
crossing in the left lower strip.

It is interesting to study the properties of the special cosmological
evolution mentioned above ($D= D_{sep}$) which separates in the upper left
strip the subset of trajectories attaining a maximum value of $s$ and then
entering the rectangle at point P  from the subset of those trajectories for
which $s$ is not bounded above and which encounter the Big Brake already in
the left upper strip. This separatrix is composed by two branches, one in
the upper left strip and a symmetrical one in the lower left strip both
having the vertical line $T=T_{3}$ as asymptote.
\begin{table}[t]
\caption{Key moments in the evolution of the tachyon universes for $k=0.2$.
Columns (1) and (2) report different pairs of values of the magnitudes $%
y_{0} $ and $x_{0}$ which are compatible with the supernovae data within 1$%
\protect\sigma $ confidence level. Columns (3), (4), (5) and (6) report
respectively the times $t_{\ast }$, $t_{BB}$ and $t_{BC}$ elapsing from the
present to the tachyonic crossing, to the attainment of the Big Brake and to
the later attainment of the Big Crunch, and the time lapse between the Big
Brake and the subsequent Big Crunch. (The values of $t_{\ast }$, $t_{BB}$
and $t_{BC}$ have been calculated assuming for the Hubble parameter the
value $H_{0}=73$ km/s/Mpc.)}
\label{Table1}
\begin{center}
\begin{tabular}{c|c|c|c|c|c}
$y_{0}$ & $x_{0}$ & $t_{\ast }\left( 10^{9}yrs\right) $ & $t_{BB}\left(
10^{9}yrs\right) $ & $t_{BC}\left( 10^{9}yrs\right) $ & $\left(
t_{BC}-t_{BB}\right) $ $\left( 10^{9}yrs\right) $ \\ \hline
$-0.90$ & $0.635$ & $0.334$ & $1.042$ & $1.412$ & $0.198$ \\
$-0.85$ & $0.845$ & $2.377$ & $3.093$ & $3.300$ & $0.207$ \\
$-0.85$ & $0.860$ & $2.438$ & $3.146$ & $3.352$ & $0.206$ \\
$-0.85$ & $0.875$ & $2.505$ & $3.206$ & $3.410$ & $0.204$ \\
$-0.80$ & $0.890$ & $6.237$ & $6.927$ & $7.135$ & $0.206$ \\
$-0.80$ & $0.905$ & $6.663$ & $7.348$ & $7.554$ & $0.206$ \\
$-0.80$ & $0.920$ & $7.197$ & $7.877$ & $8.082$ & $0.205$%
\end{tabular}%
\end{center}
\end{table}
\begin{table}[t]
\caption{As in Table \protect\ref{Table1}, for $k=0.4$.}
\label{Table2}
\begin{center}
$%
\begin{tabular}{c|c|c|c|c|c}
$y_{0}$ & $x_{0}$ & $t_{\ast }\left( 10^{9}yrs\right) $ & $t_{BB}\left(
10^{9}yrs\right) $ & $t_{BC}\left( 10^{9}yrs\right) $ & $\left(
t_{BC}-t_{BB}\right) $ $\left( 10^{9}yrs\right) $ \\ \hline
$-0.80$ & $0.710$ & $0.836$ & $1.644$ & $1.933$ & $0.289$ \\
$-0.80$ & $0.725$ & $0.841$ & $1.629$ & $1.915$ & $0.286$ \\
$-0.80$ & $0.740$ & $0.847$ & $1.616$ & $1.900$ & $0.284$ \\
$-0.75$ & $0.815$ & $2.153$ & $2.952$ & $3.247$ & $0.295$ \\
$-0.75$ & $0.830$ & $2.195$ & $2.983$ & $3.277$ & $0.294$ \\
$-0.75$ & $0.845$ & $2.242$ & $3.020$ & $3.312$ & $0.292$ \\
$-0.70$ & $0.845$ & $3.845$ & $4.635$ & $4.932$ & $0.297$ \\
$-0.70$ & $0.860$ & $3.964$ & $4.746$ & $5.043$ & $0.297$ \\
$-0.70$ & $0.875$ & $4.097$ & $4.871$ & $5.168$ & $0.297$ \\
$-0.70$ & $0.890$ & $4.247$ & $5.015$ & $5.310$ & $0.295$ \\
$-0.65$ & $0.860$ & $6.182$ & $6.959$ & $7.259$ & $0.300$ \\
$-0.65$ & $0.875$ & $6.473$ & $7.243$ & $7.540$ & $0.297$ \\
$-0.65$ & $0.890$ & $6.808$ & $7.573$ & $7.870$ & $0.297$ \\
$-0.65$ & $0.905$ & $7.204$ & $7.963$ & $8.259$ & $0.296$ \\
$-0.60$ & $0.875$ & $10.253$ & $11.016$ & $11.314$ & $0.298$ \\
$-0.60$ & $0.890$ & $11.108$ & $11.866$ & $12.163$ & $0.297$ \\
$-0.60$ & $0.905$ & $12.203$ & $12.956$ & $13.251$ & $0.295$ \\
$-0.55$ & $0.875$ & $19.517$ & $20.274$ & $20.570$ & $0.296$ \\
$-0.55$ & $0.890$ & $25.030$ & $25.782$ & $26.077$ & $0.295$%
\end{tabular}%
\ $%
\end{center}
\end{table}
\begin{table}[t]
\caption{As in Table \protect\ref{Table1}, for $k=0.6$.}
\label{Table3}
\begin{center}
$%
\begin{tabular}{c|c|c|c|c|c}
$y_{0}$ & $x_{0}$ & $t_{\ast }\left( 10^{9}yrs\right) $ & $t_{BB}\left(
10^{9}yrs\right) $ & $t_{BC}\left( 10^{9}yrs\right) $ & $\left(
t_{BC}-t_{BB}\right) $ $\left( 10^{9}yrs\right) $ \\ \hline
$-0.75$ & $0.665$ & $0.548$ & $1.369$ & $1.693$ & $0.324$ \\
$-0.70$ & $0.755$ & $1.434$ & $2.289$ & $2.624$ & $0.335$ \\
$-0.70$ & $0.770$ & $1.451$ & $2.289$ & $2.623$ & $0.334$ \\
$-0.70$ & $0.785$ & $1.469$ & $2.292$ & $2.625$ & $0.333$ \\
$-0.70$ & $0.800$ & $1.489$ & $2.299$ & $2.628$ & $0.329$ \\
$-0.65$ & $0.815$ & $2.561$ & $3.401$ & $3.740$ & $0.339$ \\
$-0.65$ & $0.830$ & $2.614$ & $3.443$ & $3.782$ & $0.339$ \\
$-0.65$ & $0.845$ & $2.671$ & $3.490$ & $3.827$ & $0.337$ \\
$-0.60$ & $0.830$ & $3.854$ & $4.692$ & $5.036$ & $0.344$ \\
$-0.60$ & $0.845$ & $3.960$ & $4.788$ & $5.131$ & $0.343$ \\
$-0.60$ & $0.860$ & $4.077$ & $4.897$ & $5.237$ & $0.340$ \\
$-0.60$ & $0.875$ & $4.206$ & $5.018$ & $5.359$ & $0.341$ \\
$-0.55$ & $0.845$ & $5.510$ & $6.336$ & $6.682$ & $0.346$ \\
$-0.55$ & $0.860$ & $5.711$ & $6.530$ & $6.875$ & $0.345$ \\
$-0.55$ & $0.875$ & $5.937$ & $6.749$ & $7.091$ & $0.342$ \\
$-0.55$ & $0.890$ & $6.194$ & $6.999$ & $7.339$ & $0.340$ \\
$-0.50$ & $0.845$ & $7.460$ & $8.281$ & $8.629$ & $0.348$ \\
$-0.50$ & $0.860$ & $7.803$ & $8.617$ & $8.963$ & $0.346$ \\
$-0.50$ & $0.875$ & $8.193$ & $9.002$ & $9.345$ & $0.343$ \\
$-0.50$ & $0.890$ & $8.645$ & $9.447$ & $9.790$ & $0.343$ \\
$-0.45$ & $0.860$ & $10.668$ & $11.478$ & $11.823$ & $0.345$ \\
$-0.45$ & $0.875$ & $11.370$ & $12.174$ & $12.519$ & $0.345$ \\
$-0.45$ & $0.890$ & $12.208$ & $13.008$ & $13.349$ & $0.341$ \\
$-0.45$ & $0.905$ & $13.237$ & $14.033$ & $14.373$ & $0.340$ \\
$-0.40$ & $0.860$ & $15.044$ & $15.851$ & $16.196$ & $0.345$ \\
$-0.40$ & $0.875$ & $16.471$ & $17.273$ & $17.615$ & $0.342$ \\
$-0.40$ & $0.890$ & $18.313$ & $19.110$ & $19.453$ & $0.343$ \\
$-0.40$ & $0.905$ & $20.838$ & $21.631$ & $21.972$ & $0.341$ \\
$-0.35$ & $0.860$ & $23.487$ & $24.291$ & $24.635$ & $0.344$ \\
$-0.35$ & $0.875$ & $27.874$ & $28.674$ & $29.016$ & $0.342$ \\
$-0.35$ & $0.890$ & $36.194$ & $36.989$ & $37.328$ & $0.339$ \\
$-0.30$ & $0.845$ & $43.469$ & $44.276$ & $44.621$ & $0.345$%
\end{tabular}%
\ $%
\end{center}
\end{table}
It encounters the Big Brake singularity at $T_{3}$ at some time moment $%
t=t_{BB}$. Now, analyzing Eq. (\ref{sdot0}) for this trajectory, we find
that in the left neighborhood of $t=t_{BB}$ the dynamical variables behave
as follows:
\begin{equation}
T=T_{3}-A_{0}(t_{BB}-t)^{2/7},  \label{sep}
\end{equation}%
\begin{equation}
s=\frac{2}{7}A_{0}(t_{BB}-t)^{-5/7},  \label{sep1}
\end{equation}%
\begin{equation}
H=\left( \frac{147\sqrt{\Lambda k(k+1)}}{4A_{0}}\right)
^{1/4}(t_{BB}-t)^{3/7},  \label{sep2}
\end{equation}%
where
\begin{equation}
A_{0}=\left( \frac{49}{12}\right) ^{1/6}\Lambda ^{-5/12}(1+k)^{-5/12}k^{1/4}.
\label{sep3}
\end{equation}%
The analysis leading to these formulae is analogous to the one which led us
to formulae (\ref{tachBB})--(\ref{HubbleBB}).

If we continue (\ref{sep})--(\ref{sep3}) beyond $t=t_{BB}$ we see that the
expansion turns into a contraction, the tachyon field $T$ starting to
decrease, while its time derivative jumps at $t=t_{BB}$ from an infinite
positive value to an infinite negative one. Thus, at $t>t_{BB}$ the universe
finds itself in the lower left strip. It leaves the Big Brake along the
asymptotic line $T=T_{3},s=-\infty $, and eventually attains the Big Crunch
singularity at the point $(T=0,s=-\sqrt{\frac{k+1}{k}})$. In the lower left
strip the separatrix separates the trajectories exiting from the Big Brake
singularity at some value $T_{BB}<T_{3}$ from those which enter the strip
from the rectangle, attain a minimum value of $s$ and end in the Big Crunch.

In summary, the expansion phase of the universe, which originated from a Big
Bang, stops at $t=t_{BB}$ and turns there into a phase of contraction
leading eventually to a Big Crunch.

\subsection{The time lapse from the Big Brake to the Big Crunch}

We give here the Tables \ref{Table1}-\ref{Table3} which, for a selection of
values of the model parameter $k$, report the times elapsing from today to
the tachyonic crossing, to the attainment of the Big Brake and to the later
attainment of the Big Crunch for some trajectories which are compatible with
SNIa data within a 1-$\sigma $ confidence level.

As we can see, although the time to the tachyonic crossing and to the Big
Brake both depend strongly on the initial data within the 1-$\sigma $ domain
selected by supernovae, the time interval between the Big Brake and the Big
Crunch do \textit{not} show the same dependence. Nevertheless, this interval
exhibits a slight increase with the model parameter $k.$

\section{Conclusions}

As shown in \textbf{II} there are tachyon cosmologies described by Eqs.(\ref%
{L})-(\ref{F2}) which are compatible with the supernovae data and which are
subject to a Big Brake in the future. In this paper we have addressed some questions
about this model:
how it behaves in the distant past, whether these
model universes can produce observational signatures today and whether they
can be continued beyond the Big Brake singularity.

Having in mind the eventual construction of fully viable models, we are
comforted by the fact that the tachyonic field has a (quasi) dust-like
behavior in the past regardless of its future evolution in the parameter
range allowed by supernovae. Present supernovae data can
hardly discriminate between the evolutions going into a de Sitter phase and
those leading to a Big Brake. However, we emphasize that the large variation
of the EoS parameter at low redshifts occurring in some of the evolutions
leading to a Big Brake might be detectable with future high precision data.

The main result of the paper is the study of the cosmological evolutions
going into a Big Brake. These evolutions can be extended beyond the Big
Brake, despite the geometry becoming singular, which apparently forbids such
a continuation. However, this singularity is a soft one, as only the second
derivative $\ddot{a}$ of the scale factor diverges at $t=t_{BB}$, while $%
\dot{a}$ does not. We need to know $\ddot{a}$ to follow the evolution of the
spacetime but we only need to know $\dot{a}$ to follow the evolution of free
particles. This means that we can not continue the evolution of the
geometry, but we can univocally continue the individual world lines of
freely falling test particles (geodesics), each of these being perfectly
regular at $t=t_{BB}$. The singularity is not experienced by any individual
freely falling particle, but makes itself felt only through the equation of
geodesic deviation, which at $t=t_{BB}$ indicates that the expansion of the
geodesic congruence turns negative from positive.

Since the geometry can be reconstructed by the knowledge of each of its
geodesics, the evolution of the universe does not stop at the Big Brake.
Once the particles have gone through the Big Brake, they will determine the
geometry anew and we can start to evolve the universe beyond the singularity. A
phase of contraction follows, leading eventually to a Big Crunch.

We have analytically and numerically analyzed the evolution of the tachyonic
universe from the Big Brake to the Big Crunch. Quite remarkably, the
numerical study showed that although the time to the tachyonic crossing and
to the Big Brake both depend strongly on the initial data chosen from the 1-$%
\sigma $ domain selected by supernovae, the time intervals between the Big
Brake and the Big Crunch do \textit{not} exhibit the same dependence\ and
they only slightly depend on the model parameter $k$ (within a factor of
two). This seems to provide an \textit{invariant timescale} for the class of
tachyonic scalar cosmologies considered, presumably related to the fact that
some information (the behavior of the higher derivatives of the scale
factor) is lost while passing through the Big Brake.

\section*{Acknowledgements}

L.\'{A}.G. wishes to thank Rachel Courtland for insightful questions. V.G.
and A.K. are grateful to Ugo Moschella for stimulating discussions. In
addition A.K. is grateful to Andrei M. Akhmeteli for illuminating comments.
Z.K. and L.\'{A}.G. were partially supported by the OTKA grant 69036; A.K.
was partially supported by RFBR grant No. 08-02-00923.

\end{document}